\documentclass[ aip, cha, jmp, amsmath,amssymb, reprint, ]{revtex4-1}

\usepackage{amssymb}
\usepackage{amsfonts}
\usepackage{amsmath}
\usepackage{amsthm}
\usepackage{epsfig}
\usepackage{color} 
\usepackage{subfigure}

\graphicspath{{./}{./Figures/}}

\begin{document}

\title{Chimera states in brain networks: empirical neural vs. modular fractal connectivity}

\author{Teresa~Chouzouris}
\affiliation{Institut f{\"u}r Theoretische Physik, Technische Universit{\"a}t Berlin, Hardenbergstra\ss{}e 36, 10623 Berlin, Germany}

\author{Iryna~Omelchenko}
\affiliation{Institut f{\"u}r Theoretische Physik, Technische Universit{\"a}t Berlin, Hardenbergstra\ss{}e 36, 10623 Berlin, Germany}

\author{Anna~Zakharova}
\affiliation{Institut f{\"u}r Theoretische Physik, Technische Universit{\"a}t Berlin, Hardenbergstra\ss{}e 36, 10623 Berlin, Germany}

\author{Jaroslav~Hlinka} 
\affiliation{Institute of Computer Science, Czech Academy of Sciences, Pod Vodarenskou vezi 2, 18207 Prague, Czech Republic}
\affiliation{National Institute of Mental Health, Topolov\'{a} 748, 250 67 Klecany, Czech Republic}

\author{Premysl~Jiruska}
\affiliation{Institute of Physiology, Czech Academy of Sciences, Videnska 1083, 14220 Prague, Czech Republic}

\author{Eckehard~Sch{\"o}ll}
\email{schoell@physik.tu-berlin.de}
\affiliation{Institut f{\"u}r Theoretische Physik, Technische Universit{\"a}t Berlin, Hardenbergstra\ss{}e 36, 10623 Berlin, Germany}

\date{\today}

\begin{abstract}
Complex spatiotemporal patterns, called chimera states, consist of coexisting coherent and incoherent domains and can be observed in networks of coupled oscillators. The interplay of synchrony and asynchrony in complex brain networks is an important aspect in studies of both brain function and disease. We analyse the collective dynamics of FitzHugh-Nagumo neurons in complex networks motivated by its potential application to epileptology and epilepsy surgery. We compare two topologies: an empirical structural neural connectivity derived from diffusion-weighted magnetic resonance imaging and a mathematically constructed network with modular fractal connectivity. We analyse the properties of chimeras and partially synchronized states, and obtain regions of their stability in the parameter planes. Furthermore, we qualitatively simulate the dynamics of epileptic seizures and study the influence of the removal of nodes on the network synchronizability, which can be useful for applications to epileptic surgery. 

\end{abstract}


\maketitle

\begin{quotation}

The investigation of synchronization in coupled oscillatory systems is an important research field in physics, biology, and technology~\cite{BLA13,PIK01,BOC06a}.
Complex mechanisms of coexistence of synchrony and asynchrony play an important role in the understanding of functionality and malfunctions of many natural and technological systems. An interesting strikingly counterintuitive example of such complex patterns are chimera states. They exhibit a hybrid structure combining domains of both synchronized and desynchronized dynamics, and were first reported for the well-known model of phase oscillators~\cite{KUR02a,ABR04}.  
In this paper, we investigate chimera states and spatiotemporal patterns in complex networks of FitzHugh-Nagumo oscillators, which model the spiking neuron dynamics.
We compare the dynamics of the networks with two types of connectivity: an empirical structural brain network topology and a simulated modular fractal topology. We find regions of existence for chimera states and other complex patterns. In the context of application to the study of epileptic seizures, we qualitatively simulate the dynamics of epileptic seizures and analyse the removal of nodes and its influence on the general synchronizability of the networks. 

\end{quotation}

\section{Introduction}

The collective behaviour in networks of oscillators is of great current interest, having numerous applications in nature and technology. Besides complete and cluster synchronization, special attention has recently been paid to chimera states where incoherent and coherent oscillations occur in spatially coexisting domains. Surprisingly, this symmetry-breaking behaviour was first discovered for identical elements and symmetric coupling configurations~\cite{KUR02a,ABR04}.
Chimera states have been studied now in a variety of oscillatory systems~\cite{SHI04,LAI09,MOT10,PAN15,SCH16b,SAK06a,SET08,MAR10,MAR10b,WOL11a,OME10a,WOL11,OME12a,PAN13,SET13,SET14,HIZ13,ZAK15b,ZAK14,YEL14,XIE14,LAI14,MAI14,MAI15,SCH14g,WOL15,LAI15,OLM15,HIZ15,OME15a,TIN12,NKO13,MAR13,LAR13,LAR15,GAM14,WIC13,SCH14a,SIE14c,ASH15,PAN15b,BOE15,SUD15,HAR16,OME16,BIC17} and discrete maps~\cite{OME11,OME12,HAG12}, with a wide range of network topologies including regular, nonlocal, nearest-neighbour, global, interacting identical subpopulations, and irregular ones. Moreover, chimera patterns are possible in higher spatial dimensions~\cite{PAN15a,MAI14}, multilayer networks~\cite{GHO16} and time varying network structures~\cite{BUS15}.

A promising application of chimera states in nature are studies of complex partially synchronised states observed in neural networks. Synchronization and desynchronization of neural activity is essential for explaining both healthy brain function and brain disorders, such as epileptic seizures~\cite{ROT14,AND16,JIR13} and Parkinson's disease. It is well known that the synchronization pattern and the related functionality depends both on the structural connectivity and on the type of local dynamics and type and strength of coupling~\cite{BRE07,HON10,DAF11,HLI12,CAB13}; making all these parameters potentially relevant for explaining synchronization alterations. For instance, during an epileptic seizure the electrical activity in the brain is excessive and synchronous, and studying chimera states can give further insight in the underlying mechanisms of the initiation or termination of epileptic seizures. Recent studies on the architecture of the neuron interconnectivity of the human and mammalian brain have shown that the connectivity of the neuronal axon network represents a hierarchical, quasi-fractal structure~\cite{KAT09,KAT12}. The existence of chimera states in networks with hierarchical connectivity (fractal) has been recently discovered~\cite{OME15,ULO16,SAW17,KRI17}.

Chimera-like states in neural networks were reported under the notion of bump states~\cite{LAI01,LAI16}. They were also observed for nonlocally coupled Hodgkin-Huxley models~\cite{SAK06a,GLA16}, FitzHugh-Nagumo~\cite{OME13,OME15}, Hindmarsh-Rose~\cite{HIZ13,HIZ16} and Integrate-and-Fire neurons~\cite{TSI17}.

In the present manuscript, motivated by studies of epileptic seizures, we provide a comparison of the transition from asynchronous behaviour to synchrony via chimera states in an empirical structural brain network derived from diffusion-weighted magnetic resonance imaging and in a mathematically constructed network with modular fractal connectivity. Our purpose is to provide insight into the nature of the real brain connectivities by studying an artificial connectivity matrix which is constructed by a well defined iterative mathematical algorithm generating a hierarchical, nested, quasi-fractal connectivity structure. To describe the dynamics of individual units, we use the paradigmatic FitzHugh-Nagumo oscillator, describing the activation and inhibition dynamics of a spiking neuron. We analyse and compare the network dynamics and complex spatiotemporal patterns in both topologies. Additionally, we analyse the influence of the removal of nodes on the network synchrony, an important aspect for the understanding of the dynamical nature of epileptic seizures~\cite{JIR14} and for improving the outcome of epilepsy surgery~\cite{SIN16a,GOO16}.

\section{The model}
\label{sec:model}

In our study, we consider a system of $N$ identical coupled FitzHugh-Nagumo oscillators:
\begin{equation}
\begin{aligned}
\varepsilon \frac{du_k}{dt} = & u_k-\frac{u_k^3}{3}-v_k\\
& +\sigma \sum\limits_{j=1}^N G_{kj} \left[b_{uu}(u_j-u_k)+b_{uv}(v_j-v_k)\right]\\
\frac{dv_k}{dt} = & u_k+a\\
& +\sigma \sum\limits_{j=1}^N G_{kj} \left[b_{vu}(u_j-u_k)+b_{vv}(v_j-v_k)\right],\\
\end{aligned}
\label{equ1}
\end{equation}
where $u$ and $v$ are the activator and inhibitor variables and all indices $k=1,...,N$ are taken modulo $N$. The parameter $\varepsilon>0$ characterizes the timescale separation, it is fixed at $\varepsilon=0.05$ throughout the paper. The coupling strength is determined by $\sigma$, and $a$ is the threshold parameter. In this study the oscillatory regime of the FitzHugh-Nagumo oscillators is considered, i.e., $|a|<1$. 

The adjacency matrix $G$ defines the topology of the network. The two investigated topologies are (i) a symmetrized empirical structural neural connectivity $G_{emp}$ obtained from diffusion-weighted magnetic resonance imaging, and (ii) a mathematically constructed topology with modular fractal connectivity $G_{mod}$. In both cases the adjacency matrix $G$ is weighted and undirected.

The local interaction scheme is characterized by a rotational coupling matrix~\cite{OME13}:
\begin{equation}
\begin{aligned} 
\mathbf{B}=\begin{pmatrix}b_{uu} & b_{uv} \\b_{vu} & b_{vv} \end{pmatrix}=\begin{pmatrix}\cos\phi & \sin\phi \\-\sin\phi & \cos\phi \end{pmatrix}~.
\end{aligned}
\label{equ2}
\end{equation}
The coupling phase $\phi$ is fixed at $\phi=\frac{\pi}{2}-0.1$, allowing information transfer between the activator and inhibitor variable of the oscillators. Recently, it was shown that the cross-couplings between activator and inhibitor in systems of identical nonlocally coupled FitzHugh-Nagumo oscillators  are essential for the observation of chimera states~\cite{OME13}.

\subsection{Empirical structural topology}
\begin{figure}[ht!]
\includegraphics[width=\linewidth]{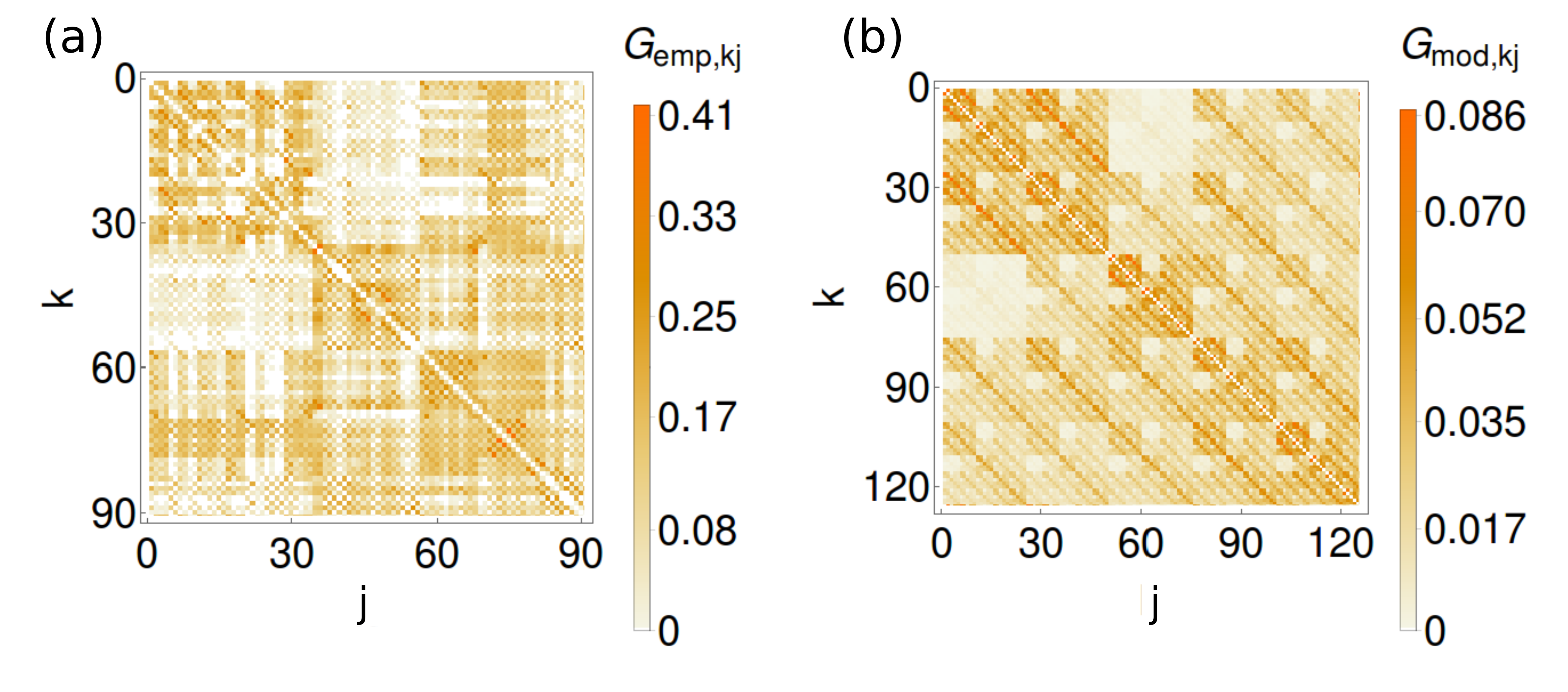}
\caption{Adjacency matrix of a network topology with (a)~empirical structural connectivity and (b)~modular fractal connectivity with the base matrix from Eq.~(\ref{equ3}).}
\label{fig1}
\end{figure}
The structural brain network of a healthy human subject shown in Fig.~\ref{fig1}(a) was obtained from diffusion-weighted magnetic resonance imaging data measured in a randomly selected member of the healthy control group that took part in the Early-Stage Schizophrenia Outcome study (ESO). For details of the original study including data acquisition parameters, see~\cite{MEL15}.
The data was analysed using probabilistic tractography~\cite{BEH07} as implemented in the FMRIB Software Library (FSL) where FMRIB stands for Functional Magnetic Resonance Imaging of the Brain (www.fmrib.ox.ac.uk/fsl/, Oxford). Note that diffusion-weighted magnetic resonance imaging is a principally 'anatomical' method, different from the well-known functional magnetic resonance imaging (fMRI), that provides indirect measurement of neuronal activity over time. The data were acquired during awake resting state with the instruction to stay as still as possible during the scanning. The anatomic network of the cortex and subcortex is constructed using Diffusion Tensor Imaging (DTI) and is divided into $90$ cortical and subcortical regions according to the Automated Anatomical Labelling (AAL) atlas~\cite{TZO02}, each region corresponding to a node in the network. Diffusion Magnetic Resonance Imaging (dMRI) provides information on the distribution of preferred diffusion direction in each volume element (voxel) of the brain, giving thus indirect information for probabilistic estimation of the trajectory of white matter fibres connecting different brain areas. Probabilistic tractography then provides for each voxel a set of $N_s$ streamlines simulating the possible white matter tracts. A coefficient $P_{ij}$ giving the connectivity probability from the $i$-th to the $j$-th region (node) can thus be introduced for all $i,j \in \{1,\dots,90\}$. It is defined as the number of probabilistic tracts connecting voxels in the i-th to voxels in the j-th area, normalized by $N_s=5000$ times the number of voxels in region $i$ -- giving the probability of a streamline entering $j$, conditional on starting in region $i$. Because of the high correlation of $P_{ij}$ and $P_{ji}$, the connectivity probability is approximated with an undirected connectivity probability given by the average of $P_{ij}$ and $P_{ji}$ for each pair of areas $(i,j)$. This way, a weighted and undirected (symmetric) adjacency matrix of size $90 \times 90$ is constructed, where each node corresponds to a cortical or subcortical area and each link is defined by the averaged connectivity probability between a pair of areas. The detailed pipeline for constructing the structural network has been adopted from a previous study focusing on the differences between healthy subjects and schizophrenia patients~\cite{CAB13}.\\

Note that while the FitzHugh-Nagumo model is a simplified model of a single neuron, it is also often used as a generic model for excitable media on a coarse-grained level. However, studies of FitzHugh-Nagumo networks often lack realistic heterogeneity of the connectivities of the network. While the full connectivity between all neurons of the human brain is not known, as a coarse approximation here we use the matrix of structural connections among 90 well-defined anatomical regions covering the main cortical and subcortical areas.

\subsection{Topology with modular fractal connectivity}
In order to provide more insight into the nature of the empirical brain connectivities, here we use an artificial connectivity matrix which has some similar features as the empirical one (Fig.~\ref{fig1}(b)). It is constructed by a well defined iterative algorithm generating a hierarchical, quasi-fractal connectivity structure from the iteration of an initial $b \times b$ base matrix $\mathbf{A}_1$, as described in \cite{KRI17}. The $b^n \times b^n$  adjacency matrix $\mathbf{G}$ for the
$n^{\text{th}}$ hierarchy level can then be formed by taking $n-1$ Kronecker
products of the initial adjacency matrix $\mathbf{A}_1$ with itself, i.e., 
\begin{equation}\label{eq:4}
\mathbf{G}= \overbrace{\mathbf{A}_1 \otimes \ldots \otimes \mathbf{A}_1}^\text{$n$ times}.
\end{equation}
In the binary case, where $\mathbf{A}_1$ contains only ones and zeros, this is essentially the 2D version of the Cantor construction of a fractal: We start with a $b \times b$  base matrix $\mathbf{A}_1$ of size $m \times m$. If we encounter a non-zero element
in the base, we substitute it with the element times the matrix $\mathbf{A}_1$, whereas a zero is replaced
by a zero matrix of size equal to the size of $\mathbf{A}_1$. 
We repeat this substitution procedure $n$ times resulting in the adjacency matrix $\mathbf{G}$ of size $m^{n-1} \times m^{n-1}$
given in Eq.~\eqref{eq:4}. This matrix $\mathbf{G}$ then defines the coupling topology, but it is no longer a circulant matrix like in a 1D ring topology, but describes a modular topology. Note that by the method used to construct the adjacency matrix $\mathbf{G}$, $\mathbf{G}$ has a constant row sum if $\mathbf{A}_1$ has a constant row sum. 

To apply this algorithm to generate a similar structure as the empirical topology in Fig.~\ref{fig1}(a), the empirical adjacency matrix $G_{emp,il}$ is divided into $5^2$ equal squares of size $m \times m$, $m=18$. The sum of links in each square determines the $5 \times 5$ elements of the base matrix $\mathbf{A}_{1}$:
\begin{equation}
\begin{aligned}
A_{1,kj}=\sum_{\substack{i= \\ (k-1)m+1}}^{k \cdot m} \sum_{\substack{l= \\ (j-1)m+1}}^{j \cdot m} G_{emp,il},
\end{aligned}
\label{equ3}
\end{equation}
with $k,j \in \{1,\dots, 5\}$. For the empirical data of Fig. 1(a) it reads explicitly:
\begin{equation}
\mathbf{A}_{1}=
\begin{bmatrix}
5.25677 & 3.22776 & 0.02343 & 1.00899 & 0.86886\\
3.22776 & 4.77906 & 0.71110 & 1.58785 & 0.68990\\
0.02343 & 0.71110 & 5.39732 & 1.27769 & 1.03968\\
1.00899 & 1.58785 & 1.27769 & 3.83577 & 1.92157\\
0.86886 & 0.68990 & 1.03968 & 1.92157 & 4.69323\\
\end{bmatrix}
\label{equ3}
\end{equation}

The third iteration of the base matrix gives the investigated modular adjacency matrix $\mathbf{G}_{mod}=\mathbf{A}_{1} \otimes \mathbf{A}_{1} \otimes \mathbf{A}_{1}$ with $N=125$ elements. The size of the base matrix $A_1$ and the number of iterations are chosen such that the mathematically constructed network topology is of comparable size and visually similar to the empirical topology. The base matrix is weighted and, therefore, a weighted modular fractal connectivity emerges.
The link weights are normalized, and the empirical and the mathematically constructed network have equal mean link weights. Furthermore, the self-coupling is set to zero. The degree distributions of the empirical and the mathematically constructed network are compared in the Appendix.

\section{The dynamical regimes}
\label{sec:The dynamical regimes}
We aim to compare possible dynamical regimes of the system Eq.~(\ref{equ1}) with the two network topologies described above. As control parameters, we will consider the threshold parameter~$a$ of the individual FitzHugh-Nagumo units, and the coupling strength~$\sigma$. Starting from random initial conditions, we provide numerical simulations of the underlying networks and observe a plethora of dynamical behaviours.
  
Together with the coexistence of coherent and incoherent domains in space, the second significant feature of chimera states is the difference of averaged oscillator frequencies. Usually, the oscillators belonging to the coherent domains have identical frequencies, and oscillators from incoherent domains are characterized by higher or lower mean frequencies. If the average time window is sufficiently large and the network is characterized by a regular topology, characteristic arc-like mean phase velocity profiles can be formed. In networks with irregular or all-to-all coupling topologies in some cases spatial reordering of the nodes is necessary to group the oscillators with equal average frequencies.
\begin{figure}[ht!]
\includegraphics[width=1\linewidth, angle=0]{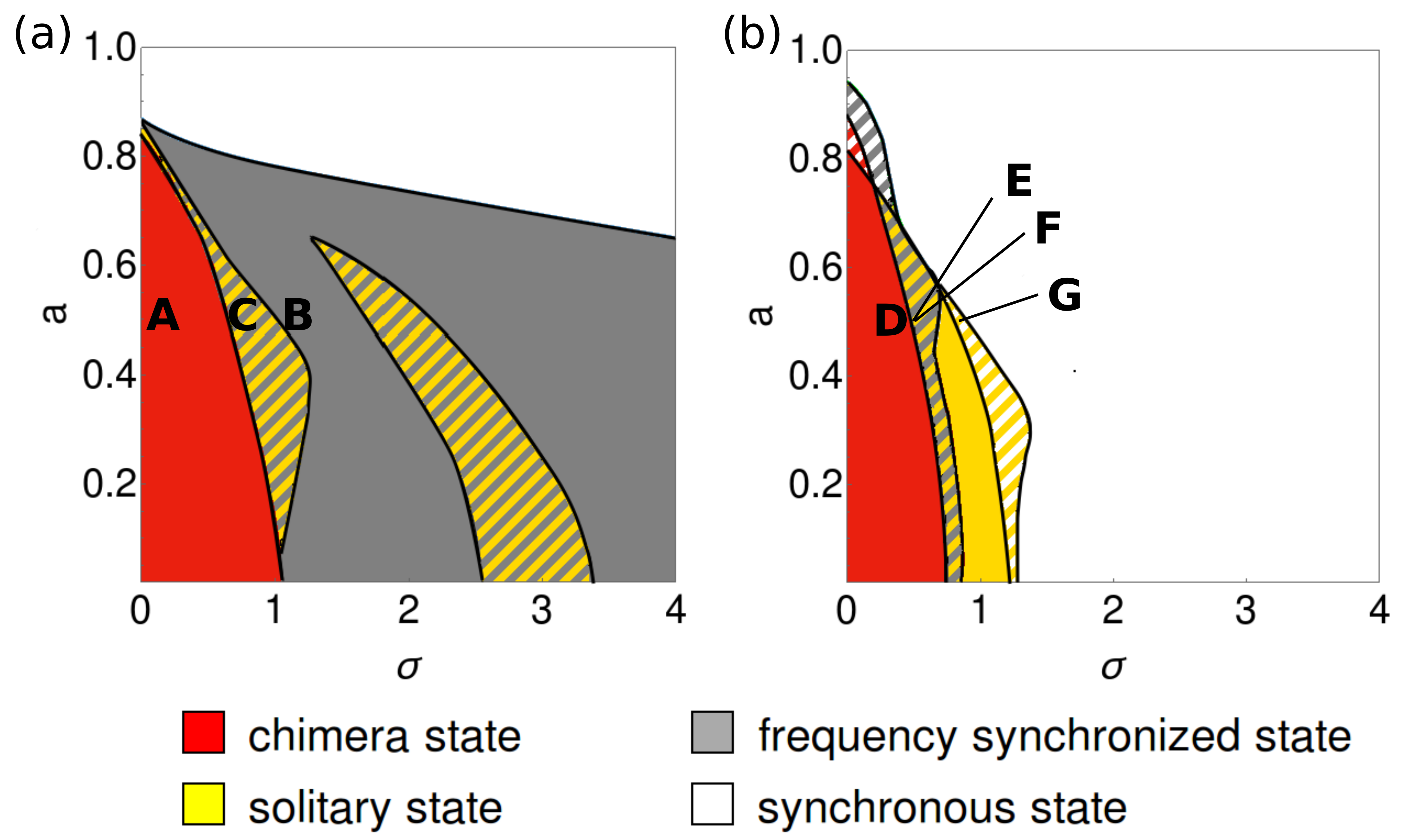}
\caption{Map of regimes in the parameter space of coupling strength~$\sigma$ and threshold parameter~$a$. (a)~Empirical structural connectivity, $N=90$, (b)~modular fractal connectivity, $N=125$. Hatched regions denote multistability of dynamical states shown with corresponding colours. The letters A, B, C, D, E, F, G mark parameter values corresponding to Figs.~\ref{fig4}(a),\ref{fig5}(a),\ref{fig6}(a),\ref{fig4}(e),\ref{fig5}(d,g),\ref{fig6}(d), respectively.}
\label{fig2}
\end{figure}
The mean phase velocities of the oscillators are calculated as $\omega_k = 2\pi M_k / \Delta T,$ $k=1,...,N,$ where $M_k$ denotes the number of complete rotations realized by the $k$th oscillator during the time $\Delta T$. As a measure for the characterization of the mean phase velocity profiles, we use the standard deviation $\Delta_{\omega} = \sqrt{(1/N) \sum_{k=1}^{N}(\omega_k - \bar{\omega})^2},$ where $\bar{\omega}=(1/N) \sum_{k=1}^N \omega_k$. Larger values of $\Delta_{\omega}$ thus correspond to more pronounced frequency differences, giving an opportunity to uncover possible chimera states.    

For the characterization of the spatial structure of the obtained patterns, we use the global Kuramoto order parameter $r=\left| \dfrac{1}{N} \sum\limits_{k=1}^N e^{i\phi_k} \right|,$ $k=1,...,N,$ where $\phi_k$ is the dynamical phase. In the uncoupled case, the geometrical phase is defined as $\tilde{\phi_k}(t)=arctan(v_k(t)/u_k(t))$. The function $t(\tilde{\phi_k})$ is calculated numerically, assigning a value of time $0<t(\tilde{\phi_k})<T$ for every value of the geometrical phase, where $T$ is the oscillation period. The dynamical phase is then defined as $\phi_k=2 \pi \cdot t(\tilde{\phi_k})/T$, which yields constant phase velocity $\dot{\phi_k}$. The global order parameter varies between $1$ and $0$, and $r=1$ corresponds to the completely synchronized state in the system. Small values denote spatially desynchronized states.
  
Fig.~\ref{fig2} depicts the numerically obtained maps of regimes for empirical and modular fractal networks of FitzHugh-Nagumo oscillators, respectively, in the plane of the coupling strength~$\sigma$ and the threshold parameter of the individual units~$a$. Note that the two systems exhibit a qualitatively similar dynamical picture with cascades of chimera states, solitary states~\cite{MAI14a}, frequency synchronised, and fully coherent states that are presented in detail in the following.  

\begin{figure*}[ht!]
\includegraphics[width=0.75\linewidth, angle=0]{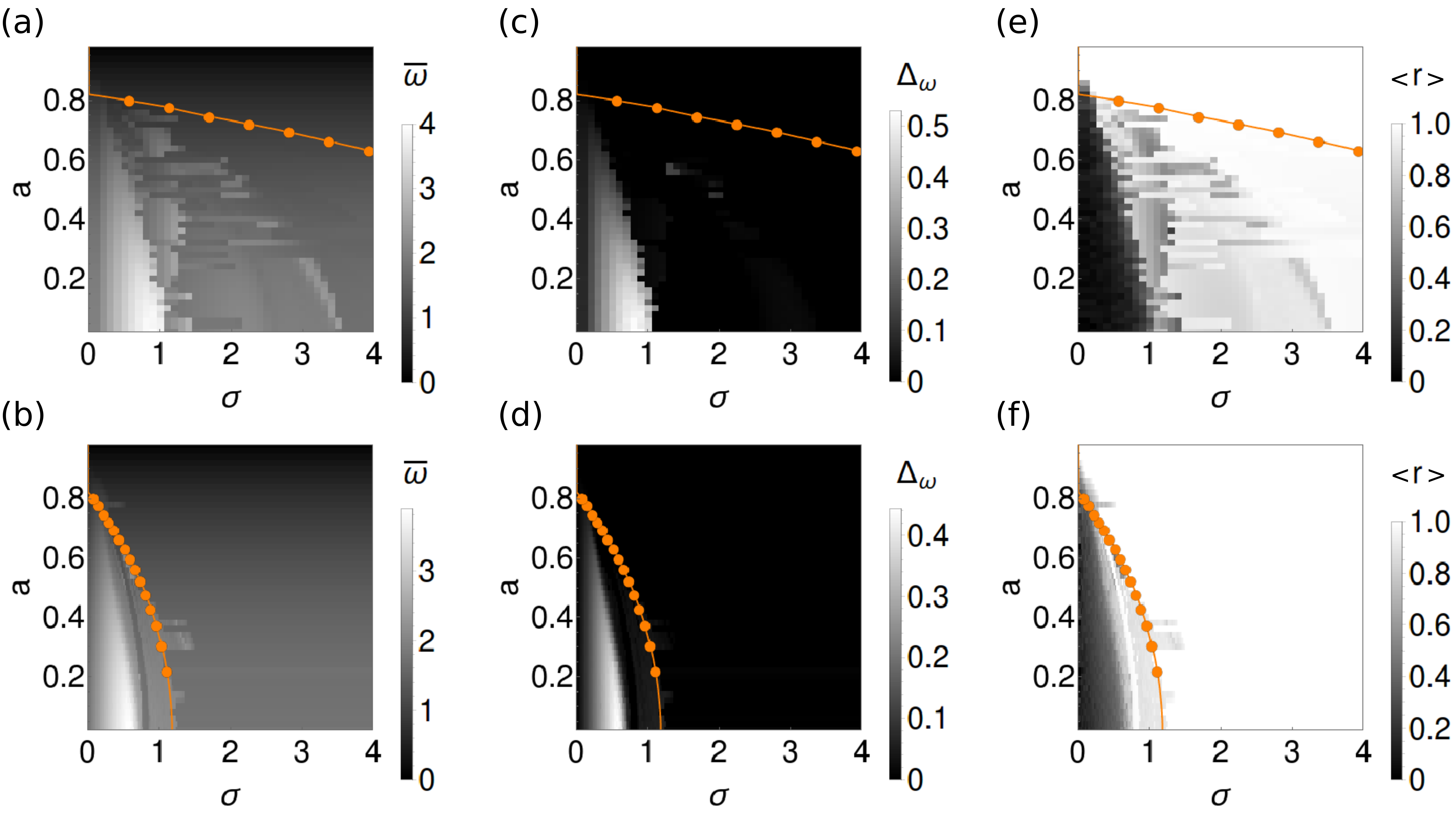}
\caption{Spatially averaged mean phase velocity $\bar{\omega}$ (left panels), standard deviation of the mean phase velocities $\Delta_{\omega}$ (middle panels), and the temporal average of the global Kuramoto order parameter~$\langle r \rangle$ (right panels) in the $(\sigma,a)$ parameter space. The orange dots depict the stability boundary of the synchronous state evaluated from the Master Stability Function; it is stable in the parameter regime above this line. (a),(c),(e)~empirical structural connectivity, $N=90$; (b),(d),(f) modular fractal connectivity, $N=125$.}
\label{fig3}
\end{figure*}

Fig.~\ref{fig3} shows the numerical evaluation of three quantities: Spatially averaged mean phase velocity $\bar{\omega}$, their standard deviation~$\Delta_{\omega}$, and the temporal average of the global Kuramoto order parameter~$\langle r \rangle$ for both types of connectivity. The initial conditions for the smallest values of the coupling strength are random. The obtained final state is used as an initial condition for the subsequent set of parameters, and so forth. Additionally, we calculate the Master Stability Function~\cite{PEC98} which defines the stability regime for the completely synchronized state. In Fig.~\ref{fig3} the orange dots show the stability boundary of the synchronous state, which is stable in the parameter regime above this line, i.e., the Master Stability Function is negative for all eigenvalues of the adjacency matrix. 

For small values of the coupling strength $\sigma$ and the threshold parameter in the approximate range $a\in(0,0.8)$ both networks exhibit chimera states (shown red in Fig.~\ref{fig2}). This is in line with earlier observations of chimera states in ring networks of FitzHugh-Nagumo oscillators with nonlocal coupling~\cite{OME13}, where chimera states are usually found for weak coupling strength and a similar range of the threshold parameter. 

\begin{figure}[hb!]
\includegraphics[width=1\linewidth]{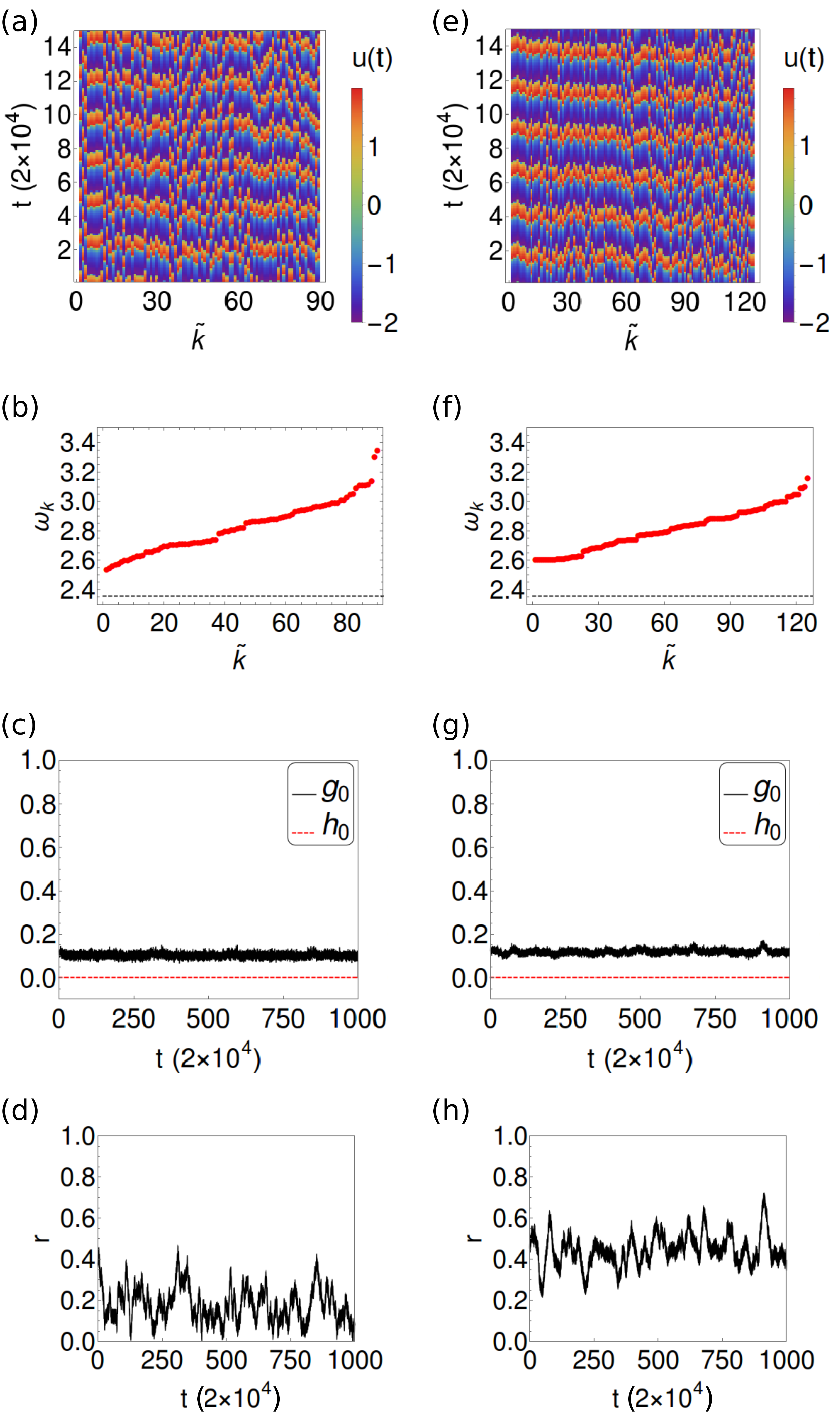}
\caption{Chimera states: (a)-(d) Network with empirical structural connectivity, $a=0.5$, $\sigma=0.2$, $N=90$; (e)-(h) Network with modular fractal connectivity, $a=0.5$, $\sigma=0.3$, $N=125$. (a),(e)~Space-time plots of $u_k$. Oscillators are reordered from lowest to highest value of $\omega_k$ (index $\widetilde{k}$); (b),(f)~Mean phase velocity profile $\omega_k$; the mean phase velocity of an uncoupled oscillator is depicted by a thin black dotted line; (c),(g)~Spatial and temporal coherence measures $g_0$ and $h_0$ vs. time; (d),(h)~Dynamics of global Kuramoto order parameter~$r$.}
\label{fig4}
\end{figure}

Examples of chimera states are shown in Fig.~\ref{fig4}, where panels (a)-(d) correspond to the network with empirical structural connectivity and parameter values marked by {\bf A} in Fig.~\ref{fig2}(a), and panels (e)-(h) correspond to the network with modular fractal connectivity with parameter values marked by {\bf D} in Fig.~\ref{fig2}(b). The upper panels~\ref{fig4}(a),~\ref{fig4}(e) in Fig.~\ref{fig4} show space-time plots for the activator variables~$u_k$. The middle panels~\ref{fig4}(b),~\ref{fig4}(f) depict the mean phase velocity profiles, where the network nodes are reordered by increasing $\omega_k$ (index $\widetilde{k}$). We observe well pronounced differences of the mean phase velocities, which is usually a prominent signature of chimera states. Due to the spatial motion of the incoherent domains, which is common for chimera states in networks of small or moderate size~\cite{OME10a}, and the two-dimensionality of the considered topologies, the mean phase velocity profile does not allow for a clear distinction of a coherent and an incoherent domain. For both networks the mean phase velocity profile is independent of the initial conditions. 
\begin{figure*}[ht!]
\includegraphics[width=0.7\linewidth]{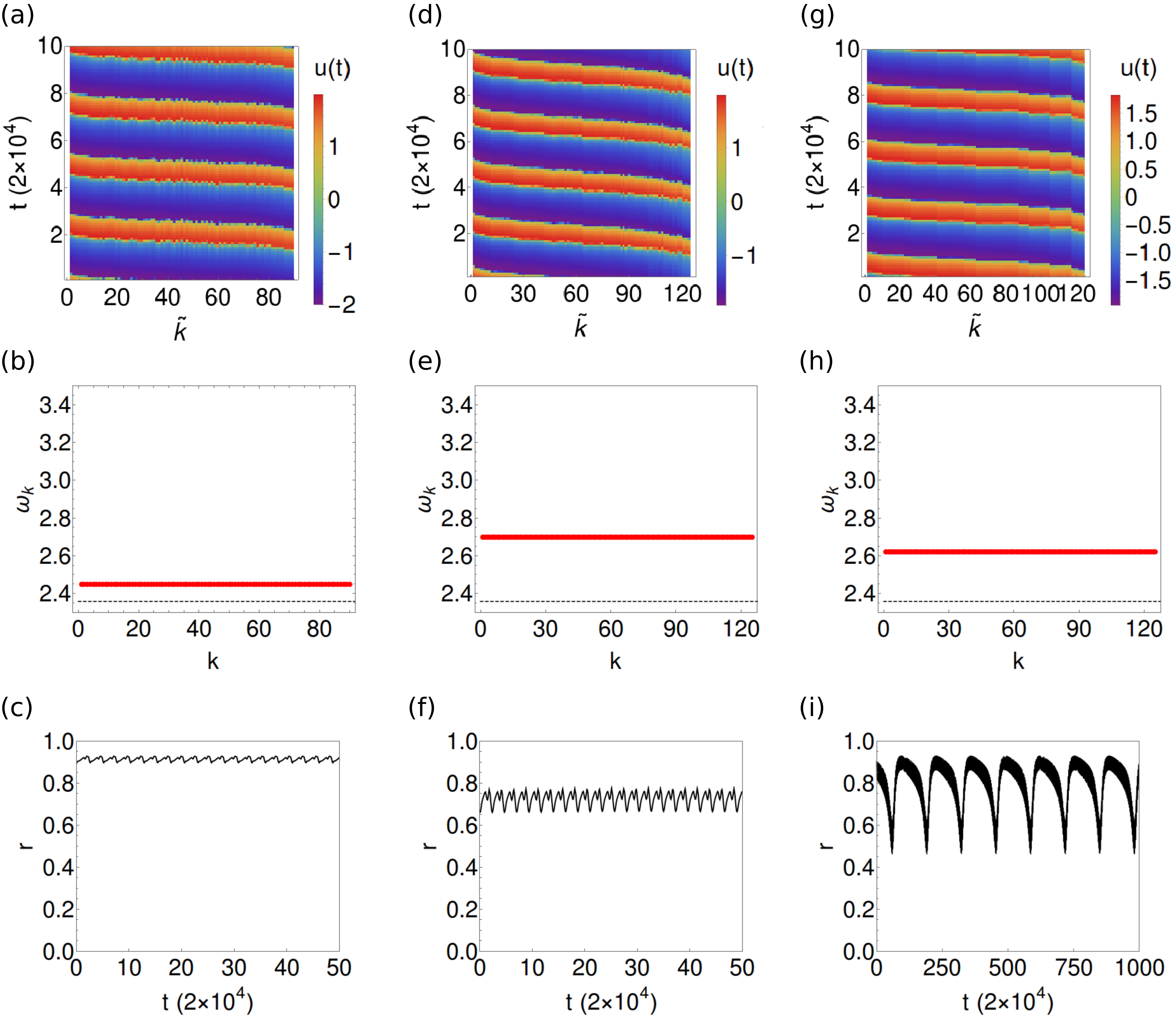}
\caption{Frequency synchronized state: (a)-(c) network with empirical structural connectivity, $a=0.5$, $\sigma=1.3$, $N=90$; (d)-(f) network with modular fractal connectivity, $a=0.5$, $\sigma=0.45$, $N=125$. (g)-(i) Breathing frequency-synchronized state: network with modular fractal connectivity, $a=0.5$, $\sigma=0.51)$, $N=125$. (a),(d),(g) Space-time plots of $u_k$. Oscillators are reordered from lowest to highest value of $u_k$ at a fixed time (index $\widetilde{k}$); (b),(e),(h) mean phase velocities $\omega_k$; the mean phase velocity of an uncoupled oscillator is shown as thin black dotted line; (c),(f),(i) dynamics of global Kuramoto order parameter~$r$.
}
\label{fig5}
\end{figure*}
The middle panels~\ref{fig4}(c),~\ref{fig4}(g) present the spatial and temporal coherence measures $g_0$ and $h_0$, respectively, introduced by Kemeth et al.~\cite{KEM16} for a general classification of chimera patterns. Corresponding to this classification scheme, the states presented in Fig.~\ref{fig4} are {\it stationary moving chimeras}. The network's temporal coherence is equal to zero and the spatial coherence is constant, indicating that the incoherent domain of the chimera states is moving in space and has a stable size. The lower panels~\ref{fig4}(d),~\ref{fig4}(h) demonstrate the dynamics of the global Kuramoto order parameter. Low values indicate the incoherence of the observed chimera patterns, and strong fluctuations of $r$ indicate the random distribution of the incoherent oscillators around the limit cycle in phase space, which changes in time.

Chimera states, which we observe both in networks with empirical structural connectivity, and networks with modular fractal connectivity, have a complex spatial structure due to the complexity of the network topologies and the absence of natural spatial ordering, in contrast to ring networks. This makes their visualization more difficult, but at the same time the spatial and temporal correlation measures clearly indicate a chimera state. 

With increasing threshold parameter~$a$ and coupling strength~$\sigma$, in both considered networks chimera states cannot be observed any more, and two other types of dynamical states occur. One of them is a frequency synchronized state shown in grey in Fig.~\ref{fig2}(a),(b). In contrast to the chimera state, now all the oscillators have equal mean phase velocities. Examples are shown in Fig.~\ref{fig5}, corresponding to parameter values marked by {\bf B} (empirical structural connectivity, Fig.~\ref{fig5}(a)-(c)), 
{\bf E} and {\bf F} (modular fractal connectivity, Fig.~\ref{fig5}(d)-(i)). The upper panels in Fig.~\ref{fig5} show space-time plots for the activator variables~$u_k$. For better visualization we reorder the network nodes corresponding to the values of $u_k$ and obtain a coherent travelling wave-like pattern. The middle panels show the mean phase velocities which are the same for all oscillators and, furthermore, they stay constant in time. The lower panels depict the dynamics of the global Kuramoto order parameter, which has large values and exhibits small periodic oscillations on the short time scale due to the amplitude dynamics, i.e., slow-fast motion on the limit cycle. Moreover, in the network with modular fractal connectivity at slightly larger coupling strength we observe a {\it breathing} frequency-synchronized state (Fig.~\ref{fig5}(g)-(i)), which exhibits stronger and slower periodic oscillations of the global Kuramoto order parameter on the long time scale due to the cyclic changes of the travelling waves speed.  

For increasing values of coupling strength~$\sigma$ and threshold parameter~$a$, we observe bistability of frequency synchronized states and {\it solitary states}~\cite{MAI14a} in both considered networks. {\it Solitary states} are characterized by frequency synchronization of the majority of oscillators while single oscillators have a different frequency. The regimes of solitary states are shown in yellow in Fig.~\ref{fig2}(a),(b); the dynamics of both networks is multistable, and solitary states coexist with frequency synchronized states (hatched regions). These patterns might also be classified as {\it weak chimera states} according to the definition introduced by Ashwin and Burylko~\cite{ASH15}, where partial frequency synchronization is the main indicator of such states.
\begin{figure}[ht!]
\includegraphics[width=1\linewidth]{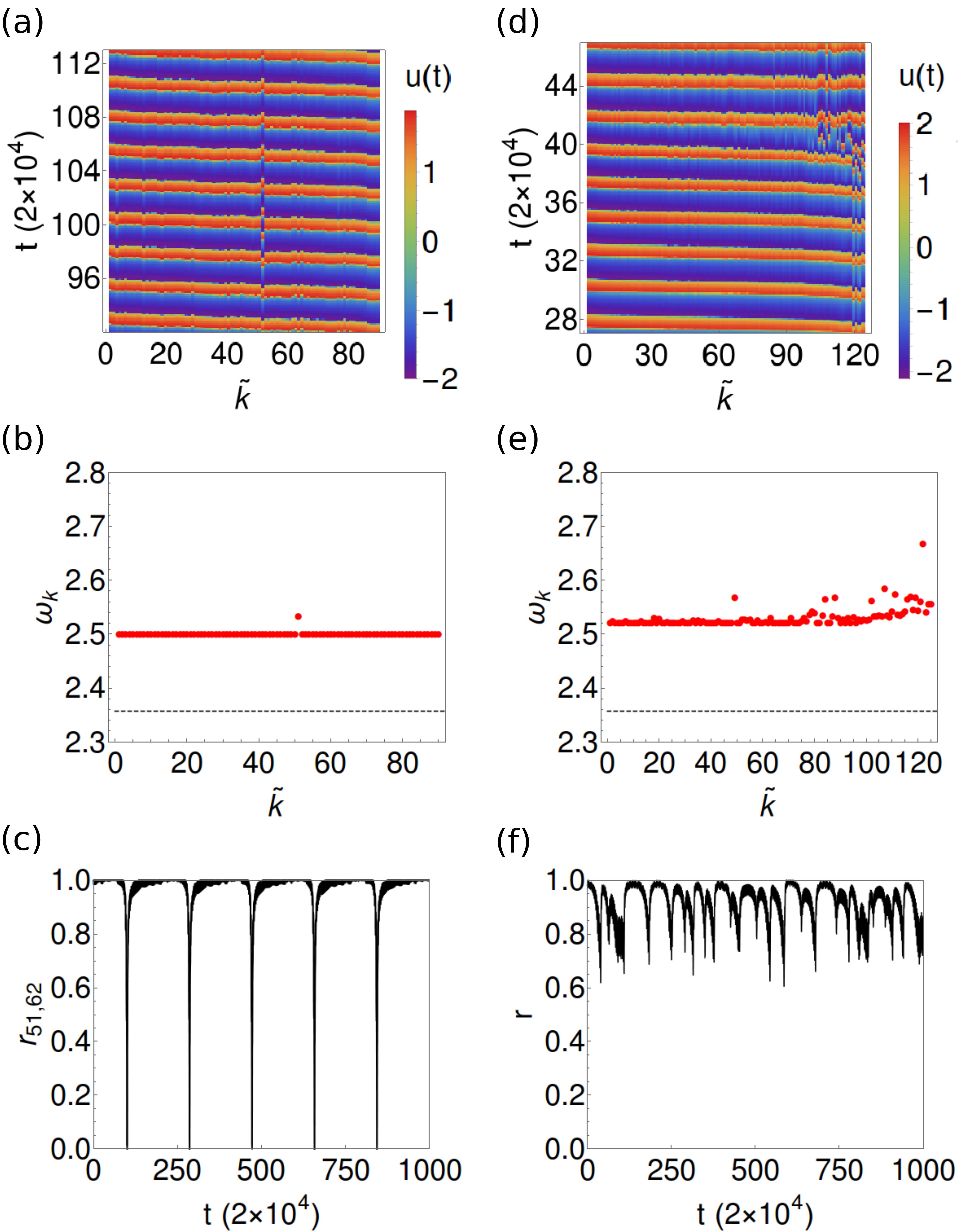}
\caption{Solitary states: (a)-(c) network with empirical structural connectivity, $a=0.5$, $\sigma=0.7$, $N=90$; (d)-(f) network with modular fractal connectivity, $a =0.5$, $\sigma=0.85$, $N=125$. (a),(d) Space-time plots of variables $u_k$, oscillators are reordered from lowest to highest value of $u_k$ at a fixed time (index $\widetilde{k}$); (b),(e) mean phase velocities $\omega_k$, the mean phase velocity of an uncoupled oscillator is shown by a thin black dotted line; (c) dynamics of Kuramoto order parameter~$r_{51,62}$ restricted to the incoherent node $\tilde{k}=51$ and a reference node $\tilde{k}=62$; (f) dynamics of global Kuramoto order parameter~$r$.
}
\label{fig6}
\end{figure}

In the network with empirical structural connectivity, we find two parameter regimes of solitary states, shown in yellow in Fig.~\ref{fig2}(a).
In the first regime, usually only one oscillator splits off, exhibiting a different mean phase velocity. This state, corresponding to parameter values marked by {\bf C} in Fig.~\ref{fig2}(a), is shown in Fig.~\ref{fig6}(a)-(c) and has one oscillator with different frequency (panel (b)). This solitary incoherent oscillator periodically, after a fixed number of periods, receives a strong input from the network and performs an additional oscillation, indicated by strong oscillations of the global order parameter (panel (c)), while the rest of the network remains frequency synchronized.
A deeper analysis shows that this solitary oscillator has the smallest clustering coefficient, and variable input from different distant parts of the network might be the reason for its incoherent behaviour. Note also the regular oscillatory dynamics of the global Kuramoto order parameter depicted in Fig.~\ref{fig6}(c). 
In the second yellow region (Fig.~\ref{fig2}(a)), for larger coupling strength~$\sigma$, we observe solitary states with three incoherent oscillators, as well as turbulent solitary states with one incoherent oscillator and non-periodic dynamics. 

In the network with modular fractal connectivity we observe similar regimes of solitary states at the transition from chimera states to complete synchronization, as well as a regime of coexistence of solitary and frequency synchronized states. An example corresponding to parameter values marked by {\bf G} in Fig.~\ref{fig2}(b), is shown in Fig.~\ref{fig6}(d)-(f). In this case, numerous solitary oscillators split off from the coherent domain. 

Hence, in both empirical and mathematically constructed networks we observe similar dynamical scenarios: for weak coupling strength we find chimera states, further increase of the coupling strength leads to a transition to complete synchronization via solitary and frequency synchronized states. For the empirical network, due to its irregular structure, larger values of coupling strength are needed, while for the modular fractal network, which due to its construction has some topological symmetries, smaller values of the coupling strength $\sigma$ are sufficient for the stabilization of the completely snychronized pattern.

\section{Simulation of epileptic seizures}
\begin{figure}[ht!]
\includegraphics[width=1\linewidth]{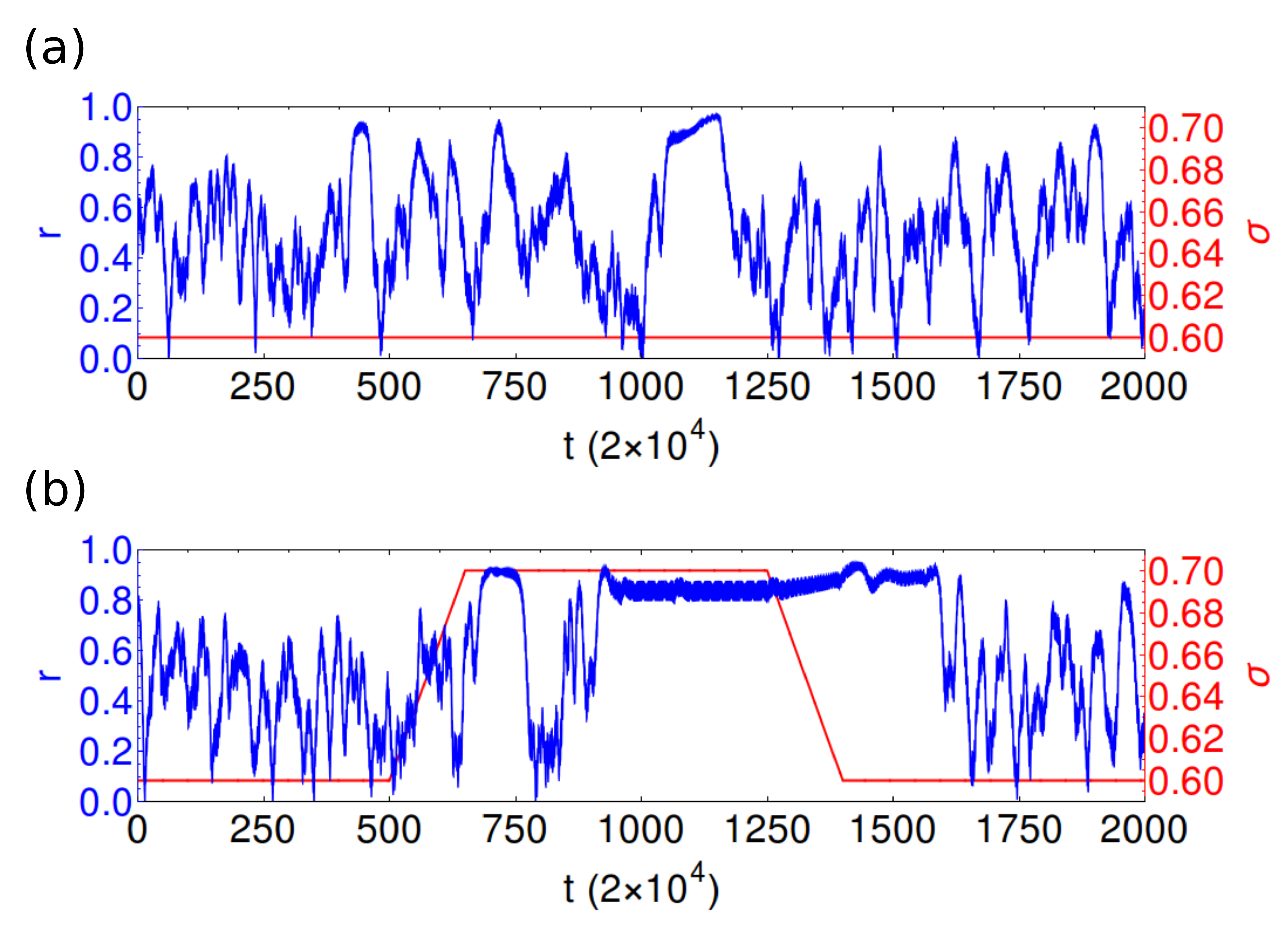}
\caption{Temporal evolution of the global Kuramoto order parameter~$r$ shown in blue for the network with empirical structural connectivity with $a=0.5$, $N=90$. The coupling strength $\sigma$ is shown in red. (a) Chimera state, constant coupling strength $\sigma=0.6$; (b) controlled dynamics: coupling strength $\sigma=0.6$ is increased to the value $\sigma=0.7$  and kept fixed for the time interval $650< t/2 \times 10^{4} \leq 1350$ , followed by decrease back to $\sigma=0.6$, this causes the transitions between the chimera state and solitary or frequency synchronized states.}
\label{fig9}
\end{figure}

Considering the empirical network, dynamics qualitatively similar to the dynamics of epileptic seizures can be computed. Spontaneous or driven high coherence events occur, indicating pathological seizure behavior. In Fig.~\ref{fig9}(a) an example of such an spontaneous event is given. The temporal evolution of the order parameter for a chimera state at fixed values $(a,\sigma)=(0.5,0.6)$ is shown. The temporal mean of the order parameter is $\langle r \rangle \approx 0,5$, and longer events of high coherence spontaneously occur. Before the high coherence event, a drop of the order parameter can be noticed, while the highest value of $r$ is obtained right before its collapse. Both effects have been observed in the study of synchrony in epileptic seizures~\cite{JIR13}, while a decrease of order preceding complete synchronization has been observed by Andrzejak et al. in a ring of nonlocally coupled phase oscillators~\cite{AND16}. The high coherence events can be controlled by the coupling strength~$\sigma$; strong coupling in the regime of chimera states increases their probability of occurrence. Furthermore, as shown in Fig.~\ref{fig9}(b), changing the coupling can induce switching between the chimera and the solitary or frequency synchronized state, which controls the pathological dynamics. It is interesting to note that the system needs some time to respond to modified coupling strength and the high synchronization event persists for some time, even after the coupling is changed back to its initial value.

\section{Node removal and synchronizability}
\label{sec:Node removal and synchronizability}
\begin{figure*}[ht!]
\includegraphics[width=0.8\linewidth ]{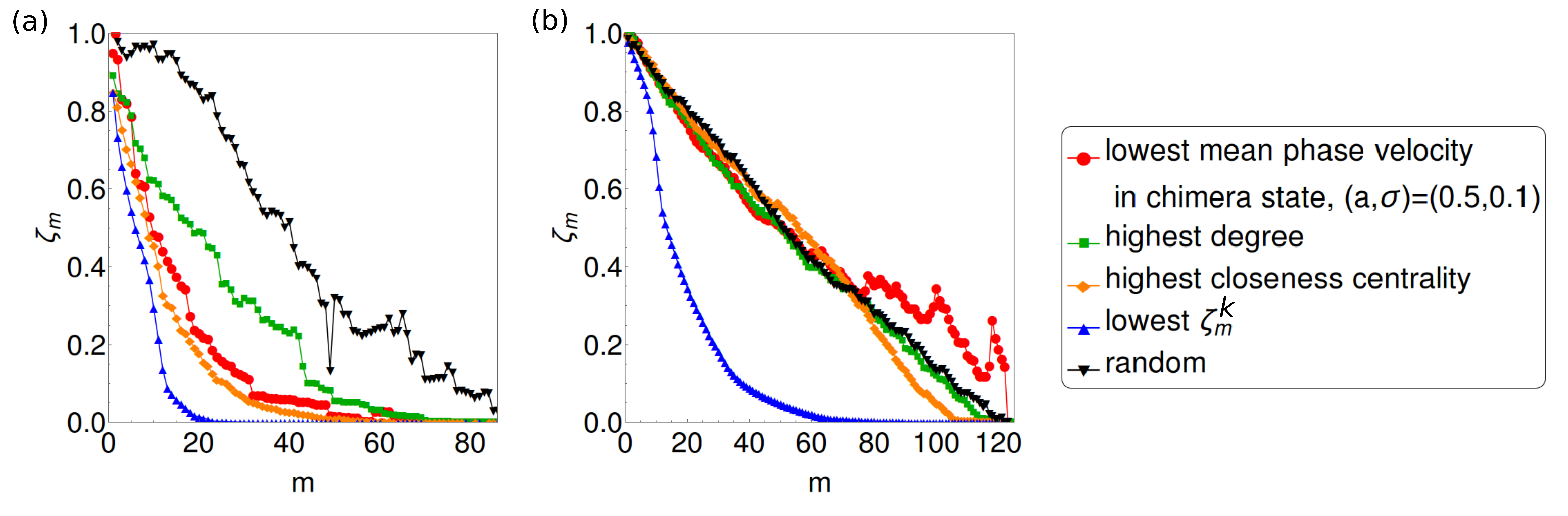}
\caption{Synchronizability index $\zeta_m$ depending on the number $m$ of progressively removed nodes. (a) Empirical structural connectivity; (b) modular fractal connectivity. All measures except the mean phase velocities are recalculated after each removal.}
\label{fig7}
\end{figure*}
In studies of epileptic seizures the focus is usually placed on possible ways to avoid the synchronization of the neural network, for this reason the analysis of the network synchronizability is of great importance. In recent years, studies have been made on the change in the seizure probability when network nodes are removed, in order incorporate the network approach in epilepsy surgery and improve its outcome~\cite{GOO16,SIN16a}. We examine the change in synchronizability in the networks with empirical structural and modular fractal connectivity, depending on dynamical and topological properties of the nodes removed. 

In our stability analysis of the synchronous solution, we apply the Master Stability approach~\cite{PEC98} to the linearized system~Eq.(\ref{equ1}):

\begin{equation}
\begin{aligned} 
\mathbf{\dot {\delta \xi}} (t)= [ D \mathbf{F} + \nu D\mathbf{B} ] \mathbf{\delta \xi}(t)~,
\end{aligned}
\label{equ4}
\end{equation}
where $\mathbf{\delta \xi}=(\mathbf{\delta u},\mathbf{\delta v})$ is a small perturbation and $D \mathbf{F}$ and $D\mathbf{B}$ are the Jacobian of the local dynamics and the interaction scheme, respectively. The Master Stability Function is the largest Lyapunov exponent $\Lambda_{max}$ expressed as a function of the complex parameter~$\nu$. The regime of $\nu$ where $\Lambda_{max}<0$, is defined as the synchronous regime $S$. 
Due to the diffusive coupling in system Eq.(\ref{equ1}) we consider the Laplacian matrix~$\mathbf{L}=\mathbf{D}-\mathbf{G}$, where $\mathbf{D}$ is the degree matrix. 
Both investigated connectivity matrices are undirected and therefore have real eigenvalues. The longitudinal Laplacian eigenvalue $\lambda_1$ is equal to zero, and all transversal Laplacian eigenvalues $\lambda_i$, $i=2,\dots ,N$, are greater than zero: $0=\lambda_1 < \lambda_2 \leq \dots \leq \lambda_N$. The synchronous solution is stable if all transversal eigenmodes lie in $S$, i.e., $-\sigma \lambda_i \in S$, $i=2,\dots ,N$. For Eq.(\ref{equ1}) $S = (-\infty,\nu_c)$, where $\nu_c<0$ is the critical value for which $\Lambda_{max}(\nu_c)=0$. Thus, the smallest transversal eigenvalue $\lambda_2$ is sufficient for determining the stability of the synchronous solution and the condition $-\sigma \lambda_2 \in S$ has to be fulfilled. The value of $\lambda_2$ is therefore used as a synchronizability index indicating the system's ability to synchronize; the closer it is to zero, the less synchronizable is the system.

Removing network nodes perturbs the topological eigenvalue spectrum of the Laplacian matrix and can cause a shift of the smallest transversal eigenvalue altering the system's synchronizability index. The synchronizability index $\zeta_m=\lambda_{2,m} / \lambda_{2,0}$, i.e., the ratio of the smallest transversal eigenvalue when $m$ nodes are removed and in the unperturbed case, measures the change in synchronizability.The incremental synchronizability index $\zeta^k_m=\lambda^k_{2,m} / \lambda_{2,m-1}$, $k \in {1,\dots,N}$ is defined as the change in synchronizability when the removal of $m-1$ nodes is increased to $m$ by removing one further node, $k \in {1,\dots,N}$. If $\zeta^k_m>1$, the smallest transversal Laplacian eigenvalue shifted to right when the $k$-th node was removed, increasing the systems ability to synchronize. If $\zeta^k_m<1$, the opposite is the case. Removing the node $k$ with the lowest value $\zeta^k_m$, thus, most effectively decreases the network's ability to synchronize.
%

\begin{figure}[ht!]
\includegraphics[width=1\linewidth]{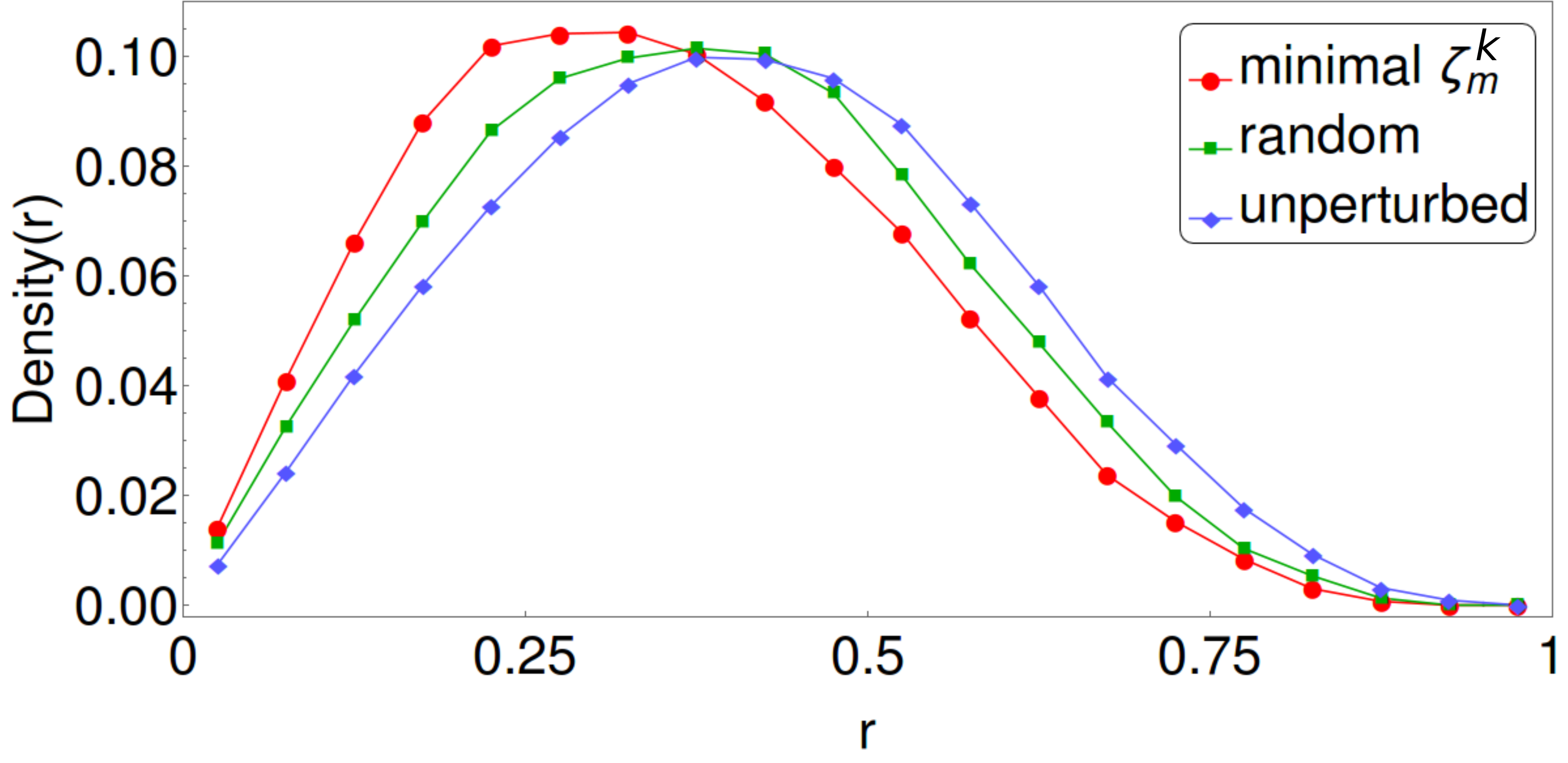}
\caption{Invariant density of the global Kuramoto order parameter~$r$ of the network with empirical structural connectivity and the chimera state with $a=0.5$, $\sigma=0.5$, when $5$ nodes with the lowest change in the synchronizability index $\zeta^k_m$ are removed in comparison to the the removal of $5$ random nodes and the unperturbed case.}
\label{fig8}
\end{figure}
In Fig.~\ref{fig7} the change of the synchronizability index $\zeta_m$ is shown, when nodes of either highest degree, or highest closeness centrality, or lowest $\zeta^k_m$, or lowest mean phase velocities in the chimera state, or randomly chosen nodes are removed progressively. All measures apart from the mean phase velocities are recalculated at every step. Removing the nodes of lowest $\zeta^k_m$ gives the optimal results. 

In the case of the network with empirical topology, removing nodes by degree or closeness centrality gives better results than random removal. Evidently, nodes that have a high influence in the network through direct and indirect links enabling quick information flow are promoting synchronization. The same result has been obtained from studies of real world networks of yeast-protein interaction, e-mail contacts and internet Autonomous Systems (AS) relationships~\cite{RES06b}. Furthermore, removing nodes of high degree most effectively reduces the mean and standard deviation of the network degrees. This finding is in line with studies reporting an increased mean and standard deviation of degrees in functional networks of epilepsy patients~\cite{CHO14c,PET14} and suggests that considering the Laplacian eigenvalue spectra will provide an explanation for this phenomenon. Counterintuitively, nodes with maximal mean phase velocities in the chimera state promote synchronous behaviour. 

In the case of the modular fractal topology the applied measures are not sensitive, with the exception of the highest eigenvalue shift. The network degree distribution is narrower and, due to the iterative construction of the topology, the node properties are not strongly distinct (see Appendix).

In the following, we discuss how knowledge on the change of the synchronizability index can be used to control the order in the network. 
In Fig.~\ref{fig8} the asymptotic invariant probability distribution 
of the global Kuramoto order parameter~$r$ of $30$ realizations with random initial conditions is depicted for fixed parameters $(a,\sigma)=(0.5,0.5)$ in the regime of chimera states. The unperturbed case is compared to cases where $5$ nodes with minimum $\zeta^k_m$, $k \in {1,\dots,N}$, $m \in {1, \dots,5}$, as well as $5$ random nodes are removed. As expected, removing nodes with minimum $\zeta^k_m$ improves the results of using random nodes and more successfully lowers the order parameter, i.e., decreases the level of synchronization.

\section{Conclusion}
\label{sec:Conclusion}

Neural networks are usually characterized by a complex topology. In order to get deeper insight into their structure and functionality,
we have compared the dynamics of a network with empirical structural neural connectivity derived from diffusion-weighted magnetic resonance imaging with a mathematically constructed network with modular fractal connectivity generated by a hierarchical iterative algorithm. 
Although the latter topology is more regular, we observe similar dynamical scenarios of complex partially synchronized states
for both networks, when we apply the FitzHugh-Nagumo model to describe the individual node dynamics.

In the plane of system parameters, we uncover the regimes of existence for chimera states which are coexisting domains of coherent and incoherent dynamics. Chimera states can be observed for weak coupling strength, and further increasing the coupling strength leads the system towards the completely coherent synchronized state. In between, we observe solitary states, where only single oscillators have different frequencies, as well as frequency-synchronized, coherent traveling wave-like states, where all oscillators have identical frequencies, but not yet synchronized amplitudes. We demonstrate that both empirical and artificially constructed networks can be characterized by qualitatively similar maps of dynamic regimes, and tuning the strength of the coupling in the network can result in the stabilization or destabilization of the completely synchronized state, as well as in partially synchronized patterns.

The knowledge of the possible dynamical regimes and transitions between them can be applied to the study of brain diseases, where synchronized behaviour is pathological, and to find possible ways to avoid it. As a prominent example we have simulated epileptic seizures where the increased coupling strength leads to pathological synchrony, initiated or terminated via chimera states, and have discussed the suppression of synchrony by the removal of nodes.  We have analysed the synchronizability of the neural networks depending on dynamical and topological properties of the nodes removed, and have shown that a topologically defined synchronizability index can be a good instrument for optimizing which nodes should be removed to decrease the synchronizability. These results from our network approach might be applied for the improvement of epilepsy surgery. 

\section*{Appendix: Degree distributions of the considered topologies}
\label{sec:Appendix}
\begin{figure}[ht!]
\includegraphics[width=1\linewidth]{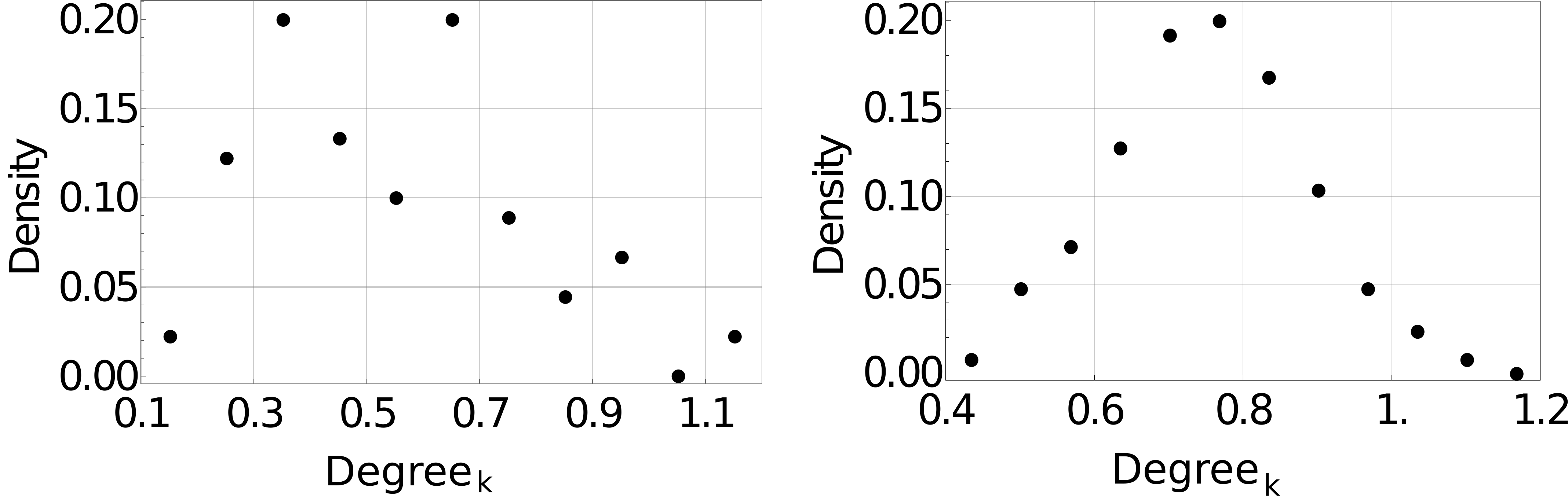}
\caption{Degree distribution of the nodes indexed by $k$ of a network topology with (a)~empirical structural connectivity and (b)~modular fractal connectivity with the base matrix from Eq.~(\ref{equ3}).}
\label{fig10}
\end{figure}

The degree distributions of the considered empirical topology and the mathematically constructed topology are depicted in figure~\ref{fig10}. Please note that since both topologies are undirected, the in-degree is equal to the out-degree.

\begin{acknowledgments}
This work was supported by Deutsche Forschungsgemeinschaft in the framework of the Collaborative Research Center SFB 910 and by grants of Ministry of Health of the Czech Republic (AZV15-29835A, AZV17-28427A) and project Nr. LO1611 with financial support from the MEYS under the NPU I program. 
We are grateful to Astero Provata, Athens, for stimulating discussion and Anton\'{i}n \v{S}koch for preparing the example structural connectivity matrix.
We dedicate this paper to the memory of John L. Hudson.
\end{acknowledgments}


\begin{thebibliography}{10}
 
 \bibitem{BLA13}
 K. Blaha, J. Lehnert, A. Keane, T. Dahms, P. H{\"o}vel, E. Sch{\"o}ll, and J.~L. Hudson, Phys.~Rev.~E {\bf 88}, 062915  (2013).
 
 \bibitem{PIK01}
 A. Pikovsky, M.~G. Rosenblum, and J. Kurths, {\em Synchronization, A Universal
   Concept in Nonlinear Sciences} (Cambridge University Press, Cambridge, 2001).
 
\bibitem{BOC06a}
 S. Boccaletti, V. Latora, Y. Moreno, M. Chavez, and D.~U. Hwang, Phys. Rep.
   {\bf 424},  175  (2006).
   
\bibitem{KUR02a}
Y.~Kuramoto and D.~Battogtokh, Nonlin. Phen. in Complex Sys.~{\bf 5}, 380~(2002).

\bibitem{ABR04}
D.~M. Abrams and S.~H. Strogatz, Phys.~Rev.~Lett. {\bf 93}, 174102~(2004).

\bibitem{SHI04}
 S.-I. Shima and Y. Kuramoto, Phys. Rev.~E {\bf 69},  036213~(2004).

\bibitem{LAI09}
C.~R. Laing, Physica~D~{\bf 238}, 1569~(2009).

\bibitem{MOT10}
A.~E. Motter, Nature Physics~{\bf 6}, 164~(2010).

\bibitem{PAN15}
 M.~J. Panaggio and D.~M. Abrams, Nonlinearity~{\bf 28}, R67~(2015). 
 
\bibitem{SCH16b}
E.~Sch{\"o}ll, Eur. Phys.~J. Spec. Top.~{\bf 225}, 891~(2016). 
 
\bibitem{SAK06a}
H. Sakaguchi, Phys. Rev.~E {\bf 73}, 031907~(2006). 
 
\bibitem{SET08}
G.~C. Sethia, A. Sen, and F.~M. Atay, Phys.~Rev.~Lett.~{\bf 100}, 144102~(2008).

\bibitem{MAR10}
E.~A. Martens, C.~R. Laing, and S.~H. Strogatz, Phys. Rev. Lett.~{\bf 104},
  044101~(2010).

\bibitem{MAR10b}
E.~A. Martens, Chaos~{\bf 20}, 043122~(2010).  
 
\bibitem{WOL11a}
M.~Wolfrum, O.~E. Omel'chenko, S.~Yanchuk, and Y.~L.~Maistrenko, Chaos~{\bf 21}, 013112~(2011).  

\bibitem{OME10a}
O.~E.~Omel'chenko, M.~Wolfrum, and Yu.~L.~Maistrenko, Phys. Rev.~E~{\bf 81}, 065201~(2010).

\bibitem{WOL11}
M.~Wolfrum and O.~E.~Omel'chenko, Phys. Rev.~E~{\bf 84}, 015201~(2011).

\bibitem{OME12a}
O.~E. Omel'chenko, M. Wolfrum, S. Yanchuk, Y.~L.~Maistrenko, and O. Sudakov, Phys. Rev.~E~{\bf 85}, 036210~(2012). 

\bibitem{PAN13}
 M.~J. Panaggio and D.~M. Abrams, Phys. Rev. Lett.~{\bf 110}, 094102~(2013).

\bibitem{SET13}
 G.~C. Sethia, A. Sen, and G.~L. Johnston, Phys. Rev.~E~{\bf 88}, 042917~(2013).
   
\bibitem{SET14}
 G.~C. Sethia and A. Sen, Phys. Rev. Lett.~{\bf 112}, 144101~(2014).
 
\bibitem{HIZ13} 
J.~Hizanidis, V.~Kanas, A.~Bezerianos, and T.~Bountis, Int. J. Bif. Chaos~{\bf 24}, 1450030~(2014).

\bibitem{ZAK15b}
A. Zakharova, M. Kapeller, and E. Sch{\"o}ll, J. Phys. Conf. Series~{\bf 727}, 012018~(2016).

\bibitem{ZAK14}
A. Zakharova, M. Kapeller, and E. Sch{\"o}ll, Phys. Rev. Lett.~{\bf 112}, 154101~(2014).  

\bibitem{YEL14}
 A. Yeldesbay, A. Pikovsky, and M. Rosenblum, Phys. Rev. Lett.~{\bf 112}, 144103~(2014). 
   
\bibitem{XIE14}
 J.~Xie, E.~Knobloch, and H.-C.~Kao, Phys. Rev.~E~{\bf90}, 022919~(2014). 
 
\bibitem{LAI14}
C.~R.~Laing, Phys.~Rev.~E~{\bf 90}, 010901~(2014).  
   
\bibitem{MAI14}
 Yu. Maistrenko, A. Vasylenko, O. Sudakov, R. Levchenko, V. Maistrenko, Int.~J. Bif.~Chaos~{\bf 24}~(8), 1440014~(2014).
 
\bibitem{MAI15}
 Yu.~Maistrenko, O. Sudakov, O.~Osiv, and V.~Maistrenko, New.~J.~Phys.~{\bf 17}, 073037~(2015). 
 
\bibitem{SCH14g}
 L.~Schmidt and K.~Krischer, Phys. Rev. Lett.~{\bf 114}, 034101~(2015).
  
\bibitem{WOL15}
M.~Wolfrum, O.~E. Omel'chenko, and J. Sieber, Chaos~{\bf 25}, 053113~(2015).
  
\bibitem{LAI15}  
C.~R.~Laing, Phys.~Rev.~E~{\bf 92}, 050904(R)~(2015).

\bibitem{OLM15}
S.~Olmi, E.~A. Martens, S.~Thutupalli, and A.~Torcini, Phys.~Rev.~E~{\bf 92}, 030901(R)~(2015).

\bibitem{HIZ15}
J.~Hizanidis, E.~Panagakou, I.~Omelchenko, E.~Sch{\"o}ll, P.~H{\"o}vel, and A.~Provata,
Phys.~Rev.~E~{\bf 92}, 012915~(2015).

\bibitem{OME15a}
I.~Omelchenko, A.~Zakharova, P.~H{\"o}vel, J.~Siebert, and E.~Sch{\"o}ll, Chaos~{\bf 25}, 083104~(2015).
 
\bibitem{TIN12}
M.~R. Tinsley, S. Nkomo, and K. Showalter, Nature Physics~{\bf 8},  662~(2012).

\bibitem{NKO13}
S.~Nkomo, M.~R.~Tinsley, K.~Showalter, Phys. Rev. Lett.~{\bf 110}, 244102~(2013).
  
\bibitem{MAR13}
 E.~A. Martens, S. Thutupalli, A. Fourri{\`e}re, and O. Hallatschek, Proc. Nat.
   Acad. Sciences~{\bf 110},  10563~(2013).
 
 \bibitem{LAR13}
 L. Larger, B. Penkovsky, and Y.~L. Maistrenko, Phys. Rev. Lett.~{\bf 111}, 054103~(2013).
   
\bibitem{LAR15} 
L.~Larger, B.~Penkovsky, and Y.~Maistrenko, Nature Comm.~{\bf 6}, 7752~(2015).
  
 \bibitem{GAM14} 
 L. V. Gambuzza, A. Buscarino, S.~Chessari, L.~Fortuna, R.~Meucci, and M.~Frasca, Phys. Rev. E~{\bf 90}, 032905~(2014).
 
 \bibitem{WIC13}
 M. Wickramasinghe and I.~Z. Kiss, PLoS ONE~{\bf 8},  e80586~(2013).
 
 \bibitem{SCH14a}
 L. Schmidt, K. Sch{\"o}nleber, K. Krischer, and V. Garcia-Morales, Chaos~{\bf 24},  013102~(2014).
   
\bibitem{ASH15} 
P.~Ashwin, and O.~Burylko, Chaos~{\bf 25}, 013106~(2015).

\bibitem{PAN15b}
M.~J.~Panaggio, D.~M.~Abrams, P.~Ashwin, and C.~R.~Laing, Phys.~Rev.~E~{\bf 93}, 012218~(2016).
  
\bibitem{BOE15}
F.~B{\"o}hm, A.~Zakharova, E.~Sch{\"o}ll, and K.~L{\"u}dge, Phys.~Rev.~E~{\bf 91}, 040901(R)~(2015).

\bibitem{HAR16} 
J.~D.~Hart, K.~Bansal, T.~E.~Murphy, and R.~Roy,  Chaos~{\bf 26}, 094801~(2016).

\bibitem{SUD15}
Y.~Suda and K.~Okuda, Phys. Rev.~E~{\bf 92}, 060901(R)~(2015).

\bibitem{BIC17}
C.~Bick, M.~Sebek, and I.~Z.~Kiss, arXiv:1705.05812~(2017).

\bibitem{OME16}
I.~Omelchenko, O.~E.~Omel'chenko, A.~Zakharova, M.~Wolfrum, and E.~Sch{\"o}ll, Phys. Rev. Lett.~{\bf 116}, 114101~(2016). 
 
\bibitem{SIE14c}
J.~Sieber, O.~E.~Omel'chenko, and M.~Wolfrum, Phys. Rev. Lett.~{\bf 112}, 054102~(2014).

\bibitem{OME11}
I. Omelchenko, Y.~L. Maistrenko, P. H{\"o}vel, and E. Sch{\"o}ll, Phys. Rev.
  Lett.~{\bf 106}, 234102~(2011). 
  
\bibitem{OME12}
I. Omelchenko, B. Riemenschneider, P. H{\"o}vel, Y.~L. Maistrenko, and E.
  Sch{\"o}ll, Phys. Rev.~E~{\bf 85}, 026212~(2012).
 
\bibitem{HAG12}
A. Hagerstrom, T.~E. Murphy, R. Roy, P. H{\"o}vel, I. Omelchenko, and E.
  Sch{\"o}ll, Nature Physics~{\bf 8},  658~(2012).
 
\bibitem{PAN15a}
M.~J. Panaggio and D.~M. Abrams, Phys. Rev.~E~{\bf 91},  022909~(2015).

\bibitem{GHO16} 
S. Ghosh, A. Kumar, A. Zakharova, and S. Jalan, Europhys. Lett.~{\bf 115}, 60005~(2016). 
  
\bibitem{BUS15}
A. Buscarino, M. Frasca, L.~V. Gambuzza, and P. H{\"o}vel, Phys. Rev.~E {\bf 91}, 2, 022817~(2015).

\bibitem{ROT14}
 A. Rothkegel, and K. Lehnertz, New J.~Phys.~{\bf 16}, 055006~(2014).  

\bibitem{AND16}
R.~G. Andrzejak, C. Rummel, F. Mormann, and K. Schindler, Sci. Rep.~{\bf 6}, 23000~(2016).

\bibitem{JIR13}
P. Jiruska, M. de Curtis, J.~G.~R. Jefferys, C.~A. Schevon, S.~J. Schiff, and K. Schindler, J.~Physiol.~{\bf 591.4}, 787~(2013).   

\bibitem{BRE07}
V. K. Jirsa, and A.~R. McIntosh, Springer Berlin Heidelberg (2007).

\bibitem{HON10}
C.~J. Honey, J.-P. Thivierge, and O. Sporns, NeuroImage~{\bf 52}, 3~(2010).

\bibitem{DAF11}
A. Daffertshofer, and B.~C.~M. van Wijk, Front Neuroinform~{\bf 5}~(2011).

\bibitem{HLI12}
J. Hlinka, and S. Coombes, Eur. J. Neurosci.~{\bf 36}, 2~(2012).

\bibitem{CAB13}
J. Cabral, H.~M. Fernandes, T.~J. Van Hartevelt, A.~C. James, M.~L. Kringelbach, and G. Deco, Chaos~{\bf 23}, 4~(2013).

\bibitem{KAT09}
P. Katsaloulis, D.~A. Verganelakis, and A. Provata, Fractals~{\bf 17}, 02, 181~(2009). 

\bibitem{KAT12}
P. Katsaloulis, A. Ghosh, A.~C. Philippe, A. Provata, and R. Deriche, Eur. Phys. J.~B~{\bf 85}, 5, 1~(2012).

\bibitem{OME15} 
I. Omelchenko, A. Provata, J.~Hizanidis, E. Sch{\"o}ll, and P. H{\"o}vel,
Phys.~Rev.~E~{\bf 91}, 022917~(2015).

\bibitem{ULO16}
S. Ulonska, I. Omelchenko, A. Zakharova, and E. Sch{\"o}ll, E., Chaos {\bf 26}, 094825~(2016). 

\bibitem{SAW17}
J. Sawicki, I. Omelchenko, A. Zakharova, and E. Sch{\"o}ll, Eur. Phys. J. Spec. Top. {\bf 226}, 1883 (2017).

\bibitem{KRI17}
S. Krishnagopal, J. Lehnert, W. Poel, A. Zakharova, and E. Sch{\"o}ll, Phil. Trans. R. Soc. A~{\bf 375}, 20160216~(2017). 

\bibitem{LAI01} 
C.~R.~Laing, C.~C.~Chow, Neural Computation~{\bf 13}, 1473 (2001).   
 
\bibitem{LAI16}
C.~R.~Laing, Frontiers Comp. Neurosc.~{\bf 10}, 53 (2016).

\bibitem{GLA16}
T.~A.~Glaze, S.~Lewis, S.~Bahar, Chaos~{\bf 26}, 083119 (2016).

\bibitem{OME13} 
I. Omelchenko, O.~E.~Omel'chenko, P. H{\"o}vel, and E. Sch{\"o}ll, 
Phys. Rev.~Lett.~{\bf 110},  224101  (2013). 

\bibitem{HIZ16}
J. Hizanidis, N.~E. Kouvaris, G. Zamora-López, A. Díaz-Guilera, and C. Antonopoulos, Sci. Rep. {\bf 6}, 19845~(2016).

\bibitem{TSI17}
N.~D. Tsigkri-DeSmedt, J. Hizanidis, E. Sch{\"o}ll, P. H{\"o}vel, and A. Provata, Eur. Phys. J. B {\bf 90}, 7, 139~(2017).
 
\bibitem{JIR14}
V.~K. Jirsa, W.~C. Stacey, P.~P. Quilichini, A.~I. Ivanov, C. Bernard, Brain {\bf 137}, 8~(2014).

\bibitem{SIN16a}
N. Sinha, J. Dauwels, M. Kaiser, S.~S. Cash, M. Brandon Westover, Y. Wang, and P.~N. Taylor, Brain {\bf 140}, 2, 319~(2016). 

\bibitem{GOO16}
M. Goodfellow, C. Rummel, E. Abela, M.~P. Richardson, K. Schindler, and J.~R. Terry, Sci. Rep. {\bf 6}, 29215~(2016).

\bibitem{MEL15}
T. Melicher, J. Horacek, J. Hlinka, F. Spaniel, J. Tintera, I. Ibrahim, P. Mikolas, T. Novak, P. Mohr, and C. Hoschl, Schizophr. Res.~{\bf 162}, 1~(2015).

\bibitem{BEH07}
T.~E.~J. Behrens, H.~Johansen Berg, S.~Jbabdi, M.~F.~S. Rushworth, and M.~W. Woolrich, NeuroImage~{\bf 34}, 1~(2007).

\bibitem{TZO02}
N. Tzourio-Mazoyer, B. Landeau, D. Papathanassiou, F. Crivello, O. Etard, N. Delcroix, B. Mazoyer, and M. Joliot, NeuroImage {\bf 15}, 1, 273~(2002). 

\bibitem{MAI14a}
Y. Maistrenko, B. Penkovsky, and M. Rosenblum, Phys. Rev. E {\bf 89}, 060901~(2014). 

\bibitem{PEC98}
L.~M. Pecora, and T.~L. Carroll, Phys. Rev. Lett. {\bf 80}, 10, 2109~(1998). 

\bibitem{KEM16}
F.~P. Kemeth, S.~W. Haugland, L. Schmidt, I.~G. Kevrekidis, and K. Krischer, Chaos {\bf 26}, 094815~(2016).

\bibitem{RES06b}
J.~G. Restrepo, E. Ott, and B.~R. Hunt, Phys. Rev. Lett. {\bf 97}, 094102~(2006). 

\bibitem{CHO14c}
F.~A. Chowdhury, W. Woldman, T.~H.~B. FitzGerald, R.~D.~C. Elwes, L. Nashef, J.~R. Terry, and M.~P. Richardson, PloS one {\bf 9}, 10, e110136~(2014).

\bibitem{PET14}
G. Petkov, M. Goodfellow, M.~P. Richardson, and J.~R. Terry, Front Neurol {\bf 5}, 261~(2014).


 
  






 \end{thebibliography}

\bibliographystyle{apsrev4-1}

\end{document}